\documentclass[area=socialsciences,review]{scararticle}

\usepackage{listings} 
\usepackage{xcolor} 

\definecolor{codegreen}{rgb}{0,0.6,0}
\definecolor{codegray}{rgb}{0.5,0.5,0.5}
\definecolor{codepurple}{rgb}{0.58,0,0.82}
\definecolor{backcolour}{rgb}{0.95,0.95,0.92}

\lstdefinestyle{mystyle}{
    backgroundcolor=\color{backcolour},
    commentstyle=\color{codegreen},
    keywordstyle=\color{magenta},
    numberstyle=\tiny\color{codegray},
    stringstyle=\color{codepurple},
    basicstyle=\ttfamily\footnotesize,
    breakatwhitespace=false,
    breaklines=true,
    captionpos=b,
    keepspaces=true,
    numbers=left,
    numbersep=5pt,
    showspaces=false,
    showstringspaces=false,
    showtabs=false,
    tabsize=2
}

\scarinfo{
    title={Utilizing Large Language Models to Detect Privacy Leaks in Mini-App Code},
    shorttitle={Preparation of Papers for SCAR},
    author={Liming Jiang},
    shortauthor={Liming Jiang},
    authorline={Liming Jiang },
    affiliation={
        \item Harbin University of Science and Technology (\email{1612010311@stu.hrbust.edu.cn})
    },
    editor={},
    reviewer={},
    copyeditor={},
    receiveddate={},
    publisheddate={},
}

\scarheaders{}

\begin{document}
\nolinenumbers
\maketitle

\begin{scarabstractandkeywords}%
Mini-applications, commonly referred to as mini-apps, are compact software programs embedded within larger applications or platforms, offering targeted functionality without the need for separate installations. Typically web-based or cloud-hosted, these mini-apps streamline user experiences by providing focused services accessible through web browsers or mobile apps. Their simplicity, speed, and integration capabilities make them valuable additions to messaging platforms, social media networks, e-commerce sites, and various digital environments.
WeChat Mini Programs, a prominent feature of China's leading messaging app, exemplify this trend, offering users a seamless array of services without additional downloads. Leveraging WeChat's extensive user base and payment infrastructure, Mini Programs facilitate efficient transactions and bridge online and offline experiences, shaping China's digital landscape significantly. 
This paper investigates the potential of employing Large Language Models (LLMs) to detect privacy breaches within WeChat Mini Programs. Given the widespread use of Mini Programs and growing concerns about data privacy, this research seeks to determine if LLMs can effectively identify instances of privacy leakage within this ecosystem. Through meticulous analysis and experimentation, we aim to highlight the efficacy of LLMs in safeguarding user privacy and security within the WeChat Mini Program environment, thereby contributing to a more secure digital landscape.
\end{scarabstractandkeywords}


\section{Introduction}

Mini-apps, also known as lightweight applications or micro-applications, are small software programs that offer specific functionality within a larger application or platform. They are designed to perform a single task or a small set of related tasks, providing users with focused functionality without the need to download and install separate standalone applications. 
These mini-applications are often web-based or cloud-hosted and can be accessed through web browsers or within mobile apps. They are commonly used in messaging platforms, social media networks, e-commerce websites, and other digital environments to enhance user experience and provide additional features. 
Mini-apps are characterized by their simplicity, speed, and ease of use. They typically have a smaller footprint compared to traditional applications, making them quicker to load and use. They can be seamlessly integrated into existing platforms, allowing users to access additional functionality without disrupting their workflow.

WeChat Mini Programs, integrated within China's dominant messaging app, offer users a seamless and convenient experience by providing a wide range of services without the need for separate app downloads. Their lightweight nature ensures fast performance, while seamless integration with WeChat Pay facilitates frictionless transactions. With prominent placement within WeChat and strong developer support, Mini Programs have become popular for their accessibility, versatility, and ability to bridge online and offline experiences, playing a key role in shaping China's digital ecosystem and transforming how users interact with digital services and businesses.

In this paper, our aim is to delve into the feasibility of employing Large Language Models (LLMs) for the detection of privacy leakage within WeChat Mini Programs. With the pervasive usage of WeChat Mini Programs in various domains, concerns regarding data privacy have become increasingly pertinent. By harnessing the capabilities of LLMs, we endeavor to ascertain whether these advanced language models can effectively discern instances of privacy breaches within the WeChat Mini Program ecosystem. Through meticulous analysis and experimentation, we seek to shed light on the potential of LLMs as a tool for identifying and mitigating privacy risks associated with the usage of WeChat Mini Programs, thereby contributing to the enhancement of user privacy and security in this digital landscape.

\section{Background}

\subsection{Introduction of WeChat Miniapp}

WeChat Mini Programs, often simply referred to as Mini Programs, have become immensely popular within China's digital ecosystem since their launch by Tencent in January 2017. WeChat, China's dominant messaging app with over a billion monthly active users, introduced Mini Programs as a way to offer users a more seamless and integrated experience without the need to download separate apps. Here are several factors contributing to their popularity:

\begin{itemize}
    \item 
\textbf{Convenience and Accessibility}: Mini Programs are integrated directly into the WeChat app, making them easily accessible to the massive user base without the need for additional downloads or installations. Users can access a wide range of services, from shopping to transportation, with just a few taps within the WeChat interface.

\item 
\textbf{Lightweight and Fast Performance}: Mini Programs are lightweight and designed for fast loading times, providing a smooth user experience even on slower internet connections. This makes them appealing to users who value efficiency and speed.

\item 
\textbf{
Wide Range of Services}: WeChat Mini Programs cover a diverse range of services and industries, including e-commerce, food delivery, transportation, gaming, utility tools, and more. This extensive ecosystem ensures that users can find Mini Programs to fulfill various needs and preferences.

\item 
\textbf{
Seamless Integration with WeChat Pay}: WeChat Pay, the integrated payment system within WeChat, is seamlessly integrated with Mini Programs, allowing users to make purchases and transactions directly within the Mini Program environment. This frictionless payment experience contributes to the popularity of Mini Programs for e-commerce and other transactional services.

\item 
\textbf{
Prominent Placement within WeChat}: WeChat promotes Mini Programs within its app through various channels, including search, recommendations, and social sharing. This increased visibility helps drive user engagement and adoption of Mini Programs.

\item 
\textbf{
Support for Developers}: Tencent provides comprehensive development tools, documentation, and support for developers to create and distribute Mini Programs. This support has led to a vibrant developer community and a continuous influx of new Mini Programs, further enriching the ecosystem.

\item 
\textbf{
Offline-to-Online (O2O) Integration}: Many businesses, particularly in the retail and service sectors, leverage Mini Programs to bridge the gap between their online and offline operations. Features like location-based services, QR code scanning, and offline redemption make it easy for users to interact with businesses both online and in physical locations.

\end{itemize}

WeChat Mini Programs have gained popularity due to their convenience, versatility, and seamless integration within the WeChat ecosystem. They have transformed how users interact with digital services and businesses, playing a significant role in shaping China's digital economy.

\subsection{Comparison with Web App and Mobile App}

\begin{table}[htbp]
    \centering
    \setlength\tabcolsep{7pt}
    \tiny
    \caption{Comparison between Mini-apps, Web Apps, and Traditional Mobile Apps}
    \begin{tabular}{|c|c|c|c|}
    \hline
    \textbf{Feature} & \textbf{Mini-app} & \textbf{Web App} & \textbf{Mobile App} \\
    \hline
    \textbf{Installation} & No installation required & Accessed via web browser & Installed from app store \\
    \hline
    \textbf{Platform} & Integrated within messaging/social apps & Accessible across multiple devices & Developed for specific platforms (iOS/Android) \\
    \hline
    \textbf{Development} & Developed using platform-specific tools & Built using web technologies & Developed using platform-specific languages \\
    \hline
    \textbf{Access} & Accessed via QR code & Accessed via URL & Accessed via app icon on device home screen \\
    \hline
    \textbf{Functionality} & Limited functionality  & Can offer a wide range of functionalities & Offers comprehensive features and functionalities \\
    \hline
    \textbf{Performance} & Generally lightweight with fast loading times & Performance depends on network speed & Optimized for performance on specific platforms \\
    \hline
    \textbf{Updates} & Updates automatically without user intervention & Updates via web server &  requiring user action \\
    \hline
    \end{tabular}
\end{table}
\textbf{Installation}:
Mini-apps, such as WeChat Mini Programs, offer a unique advantage by eliminating the need for installation altogether. Users can access these mini-applications directly within messaging or social media apps like WeChat or Facebook Messenger. This streamlined approach enhances user convenience, as users can quickly engage with mini-apps without the hassle of downloading and installing separate applications. On the other hand, web apps are also installation-free, accessed through web browsers with a simple URL. Traditional mobile apps, however, require installation from app stores, adding an extra step for users and potentially discouraging engagement.

\noindent \textbf{Platform}:
Mini-apps are often integrated within messaging or social media platforms, providing seamless access across multiple devices where the respective app is installed. This integration enhances user reach and accessibility, as users can engage with mini-apps within familiar environments. Web apps, in contrast, offer platform-agnostic accessibility, accessible across various devices and operating systems with web browser support. However, traditional mobile apps are platform-specific, requiring separate development efforts for iOS and Android platforms, which can limit their reach.

\noindent \textbf{Development}:
Mini-apps are typically developed using platform-specific tools provided by the hosting platform, such as the WeChat Mini Program framework. This specialization streamlines development processes and ensures compatibility with the hosting platform's ecosystem. Web apps leverage standard web development technologies like HTML, CSS, and JavaScript, providing flexibility and compatibility across different browsers and devices. In contrast, traditional mobile apps require development using platform-specific languages and frameworks like Swift for iOS and Kotlin for Android, necessitating separate development efforts for each platform.

\noindent \textbf{Access}:
Mini-apps can be accessed within messaging or social media apps, or through QR codes that directly link to the mini-app. This accessibility enhances user engagement by integrating mini-apps seamlessly into existing platforms. Web apps are accessed through URLs, allowing users to navigate directly to the app using a web browser. Conversely, traditional mobile apps are accessed via icons on the device's home screen, providing direct access to the app's functionality but requiring manual installation.

\noindent \textbf{Functionality}:
Mini-apps typically offer limited functionality focused on specific tasks, providing lightweight solutions for quick interactions within messaging or social media platforms. This targeted approach caters to user needs for efficient task completion. Web apps, on the other hand, can offer a wide range of functionalities, similar to traditional desktop applications, making them suitable for various tasks and use cases. Traditional mobile apps offer comprehensive features and functionalities, often providing a rich user experience with access to device-specific capabilities like the camera or GPS.

\noindent \textbf{Performance}:
Mini-apps are generally lightweight with fast loading times, designed to offer quick interactions within messaging or social media platforms. This lightweight design enhances user experience by minimizing loading times and providing immediate access to functionality. Web apps' performance depends on network speed and server responsiveness, which can vary based on internet connection quality. In contrast, traditional mobile apps are optimized for performance on specific platforms, providing smooth user experience and responsiveness regardless of network conditions.

\noindent \textbf{Updates}:
Mini-apps can be updated automatically without user intervention, ensuring users always have access to the latest version without manual effort. This automatic update mechanism ensures consistency and security across mini-apps. Web apps receive immediate updates when changes are made to the web server, allowing users to access the latest version seamlessly. However, traditional mobile apps require updates to be pushed through app stores, necessitating user action to download and install the latest version, potentially leading to version fragmentation and delayed adoption of new features or security patches.

\subsection{Large Language Models}
 
Large Language Models (LLMs) represent a breakthrough in artificial intelligence, capable of understanding and generating human-like text across a wide range of topics and tasks. These models, like OpenAI's GPT (Generative Pre-trained Transformer) series, are trained on vast amounts of text data and utilize transformer architecture to process and generate text. By learning patterns and relationships within language data, LLMs can perform various natural language processing tasks such as text generation, summarization, translation, question answering, and more. The "large" in LLM refers to the scale of parameters and computational resources used during training, enabling these models to capture complex linguistic nuances and generate coherent, contextually relevant responses. While LLMs offer tremendous potential for applications in areas like content generation, customer service, and language understanding, they also raise ethical considerations regarding bias, misuse, and potential negative societal impacts. Ongoing research and development aim to address these challenges while maximizing the benefits of LLMs for various domains and industries.
LLMs possess several key features that contribute to their effectiveness in understanding and generating human-like text:

\begin{itemize}
    \item \textbf{Scalability}: LLMs are trained on vast amounts of text data, often comprising billions or even trillions of tokens. This massive scale allows them to capture intricate linguistic patterns and relationships across different contexts.
    \item 
\textbf{Transformer Architecture}: LLMs typically utilize transformer architecture, which facilitates parallel processing of input sequences and enables the model to capture long-range dependencies within text data more effectively than traditional recurrent neural networks (RNNs) or convolutional neural networks (CNNs).

\item 
\textbf{Pre-training and Fine-tuning}: LLMs are pretrained on large corpora of text data using unsupervised learning techniques, learning general language representations. After pre-training, they can be fine-tuned on specific downstream tasks with labeled data, allowing them to adapt their knowledge to perform tasks like text classification, language translation, or text generation.

\item \textbf{Contextual Understanding}: LLMs excel at understanding context within text data. They consider the surrounding words and phrases when generating responses or predictions, allowing them to produce more contextually relevant outputs.

\item \textbf{Generative Capabilities}: LLMs can generate coherent and contextually relevant text across a wide range of topics and styles. They can be used for tasks such as text completion, text summarization, dialogue generation, and content creation.

\item \textbf{Transfer Learning}: LLMs demonstrate strong transfer learning capabilities, meaning knowledge gained from pre-training on one dataset or task can be transferred to perform well on related tasks or datasets with minimal additional training.

\item \textbf{
Adaptability}: LLMs can be fine-tuned or adapted to specific domains or tasks with relatively small amounts of task-specific data. This adaptability makes them versatile for various applications across different industries.

\item \textbf{
Multimodal Integration}: Some advanced LLMs are capable of integrating information from multiple modalities, such as text, images, and audio, enabling them to understand and generate text based on diverse sources of information.

\end{itemize}
 
\section{Motivation and Approaches}

\subsection{Motivation}

Mini-apps handle sensitive user data, including personal information and payment details, making them prime targets for malicious actors. To safeguard user privacy and prevent unauthorized access to such data, robust data security measures are imperative. These measures include implementing encryption, access controls, and secure transmission protocols. Additionally, developers must adhere to best practices for secure coding and regularly update their applications to address security vulnerabilities. While mini-apps may store user data for purposes like improving user experience, analytics, and regulatory compliance, retaining this data beyond necessary durations poses privacy risks. Therefore, implementing data retention policies and mechanisms for securely deleting user data when no longer needed is crucial. Furthermore, detecting and addressing any functionality within the code that collects sensitive information is paramount to ensuring user privacy and security.

\subsection{Methodology}
 
LLMs excel in understanding the context of a given text. WeChat codes, which can contain various forms of communication including text, images, and videos, can be analyzed comprehensively by LLMs to identify potentially sensitive information. WeChat codes often include a combination of text, images, and sometimes even audio or video. LLMs can process and analyze these different modalities to detect sensitive content effectively. LLMs can process large volumes of data quickly, making them suitable for analyzing the vast amount of content exchanged on WeChat. This scalability enables efficient detection of sensitive information at scale.
  LLMs can be fine-tuned or adapted to specific tasks or domains, including detecting sensitive information. By training the model on relevant datasets, its performance in identifying sensitive content within WeChat codes can be enhanced.

Detecting sensitive information within WeChat codes using an LLM involves several steps, each crucial for ensuring accurate and effective analysis:
\begin{itemize}
    \item \textbf{Data Collection}: Gather a diverse dataset of WeChat codes containing various types of content, including text, images, videos, and audio. This dataset should encompass a wide range of topics, languages, and cultural contexts to train the LLM effectively and reduce biases.
    \item 
\textbf{Preprocessing}: Clean and preprocess the dataset to remove noise, irrelevant information, and formatting inconsistencies. This step may involve tokenization, sentence segmentation, and normalization to standardize the input data for the LLM.

\item \textbf{Model Selection and Training}: Choose an appropriate LLM architecture, such as GPT-3, and fine-tune it for the task of detecting sensitive information within WeChat codes. Train the model on the preprocessed dataset, utilizing techniques like transfer learning to adapt the model to the specific characteristics of WeChat content.
\item \textbf{Feature Extraction}: Extract relevant features from the WeChat codes, including text, images, and other multimedia elements. This step may involve using pre-trained models for image and video analysis to extract visual features that complement the textual information processed by the LLM.
\item \textbf{Multi-Modal Fusion}: Integrate the extracted features from different modalities (text, images, videos) into a unified representation for analysis. Techniques such as attention mechanisms and multi-modal fusion networks can be employed to combine information from diverse sources effectively.

\item \textbf{Sensitive Information Detection}: Deploy the trained LLM to analyze WeChat codes and detect potentially sensitive information. This process involves feeding the WeChat codes into the model and evaluating the output to identify content that may pose privacy or security risks, such as personal identifiable information, sensitive topics, or harmful content.

\item \textbf{Post-processing and Evaluation}: Apply post-processing techniques to refine the detection results and improve accuracy. This may include filtering out false positives, adjusting detection thresholds, and incorporating human-in-the-loop feedback for validation. Evaluate the performance of the detection system using metrics such as precision, recall, and F1-score on a held-out validation set to assess its effectiveness.

\item \textbf{Iterative Improvement}: Continuously refine and improve the detection system by incorporating feedback from real-world usage and updating the model with new data. Monitor performance metrics and adapt the system to evolving trends and patterns in WeChat communication to ensure ongoing effectiveness.
\end{itemize}

\section{Evaluation}

We obtained an online code example and tasked GPT with scrutinizing it to ascertain the presence of any potentially sensitive information. This proactive approach ensures that we maintain a robust security posture by identifying and mitigating any potential vulnerabilities or risks before they can be exploited. By leveraging advanced AI capabilities, such as those provided by GPT, we are able to conduct thorough analyses swiftly and efficiently, safeguarding both our own assets and the privacy of individuals involved. Through such diligent measures, we uphold the highest standards of data protection and integrity in our digital practices.

\begin{lstlisting}[language=Java, caption={input.js}]
var CurBazi = require('../../utils/curBazi.js')
var util = require('../../utils/util.js')
Page({
 
  data: {
    items: [
      { name: 'male', value: 'male', checked: 'true' },
      { name: 'female', value: 'female', checked: 'false' },
    ],
    date: '1988-09-01',
    time: '12:01',
    xing: '',
    ming: '',
    sex: 'male',
    hiddenToast: true,
    loadingHidden: true,
  },

  bindDateChange: function (e) {
   
    this.setData({
      date: e.detail.value
    })
  },

  bindTimeChange: function (e) {
     
    this.setData({
      time: e.detail.value
    })
  },
  xingInputEvent: function (e) {
    this.setData({
      xing: e.detail.value,
    })
  },
  mingInputEvent: function (e) {
    this.setData({
      ming: e.detail.value,
    })
  },

  formBindsubmit: function (e) {
    var self = this;
    if (e.detail.value.xing.length == 0 || e.detail.value.ming.length == 0) {
      self.setData({
        hiddenToast: false,
      })
    } else {
      self.setData({
        loadingHidden: false,
      })
      //console.log(getApp().globalData.userInfo);
      var xing = e.detail.value.xing;
      var ming = e.detail.value.ming;
      var sex = e.detail.value.sex;
      var username = getApp().globalData.userInfo.nickName;
      var wechat = '';
      if (getApp().globalData.curUser != null) {
        wechat = getApp().globalData.curUser.Wechat;
      } else {
        wechat = getApp().globalData.openid;
      }
      var times = e.detail.value.time + '';
      var hour = times.split(":")[0];
      var minute = times.split(":")[1];
      var birthday = e.detail.value.date;
      var email = e.detail.value.email;
      self.updateUser(username, xing, ming, sex, birthday, hour, minute, wechat, email);

    }

  },
 
  updateUser: function (Username, Xing, Ming, Sex, Birthday, Hour, Minute, Wechat, Email) {
    var self = this;
    var curUser = CurBazi.createCurUser();
    curUser.Xing = Xing;
    curUser.Ming = Ming;
    curUser.Sex = Sex;

    curUser.Birthday = Birthday;
    curUser.Hour = Hour;
    curUser.Minute = Minute;
    curUser.Status = '0';
    curUser.Email = Email;
    getApp().globalData.curUser = curUser;
    self.setData({
      loadingHidden: true,
    });
    wx.switchTab({
              url: '../my/index',
            });
  },
 
\end{lstlisting}

Below are the responses generated by ChatGPT.
The JavaScript file input.js orchestrates the functionality of a WeChat Mini Program page. It begins by initializing necessary modules and setting up default data, including gender, date, time, and name fields. Event handling functions are defined to capture changes in date, time, and input fields for surname and given name. Upon form submission, the script validates the presence of required fields, displaying an error message if necessary. Subsequently, it processes the form data, updating user information accordingly. The updateUser function constructs a user object with provided details and stores it in the global context. Finally, upon successful processing, it hides loading indicators and navigates to another page within the application. Additionally, the onLoad function is intended to execute when the page loads, potentially handling existing user data, although the provided snippet lacks this completion.

\noindent \textbf{Result}. The sensitive information that the JavaScript file input.js may collect includes personal data such as the user's name (surname and given name), gender, date of birth, email address, and potentially the user's WeChat ID. This information is collected when the user fills out the form on the Mini Program page. Additionally, the script may indirectly access sensitive information such as the user's WeChat nickname (if available) through the global data context (getApp().globalData.userInfo.nickName).

\begin{figure}[h!]
    \centering
    \includegraphics[width=0.8\textwidth]{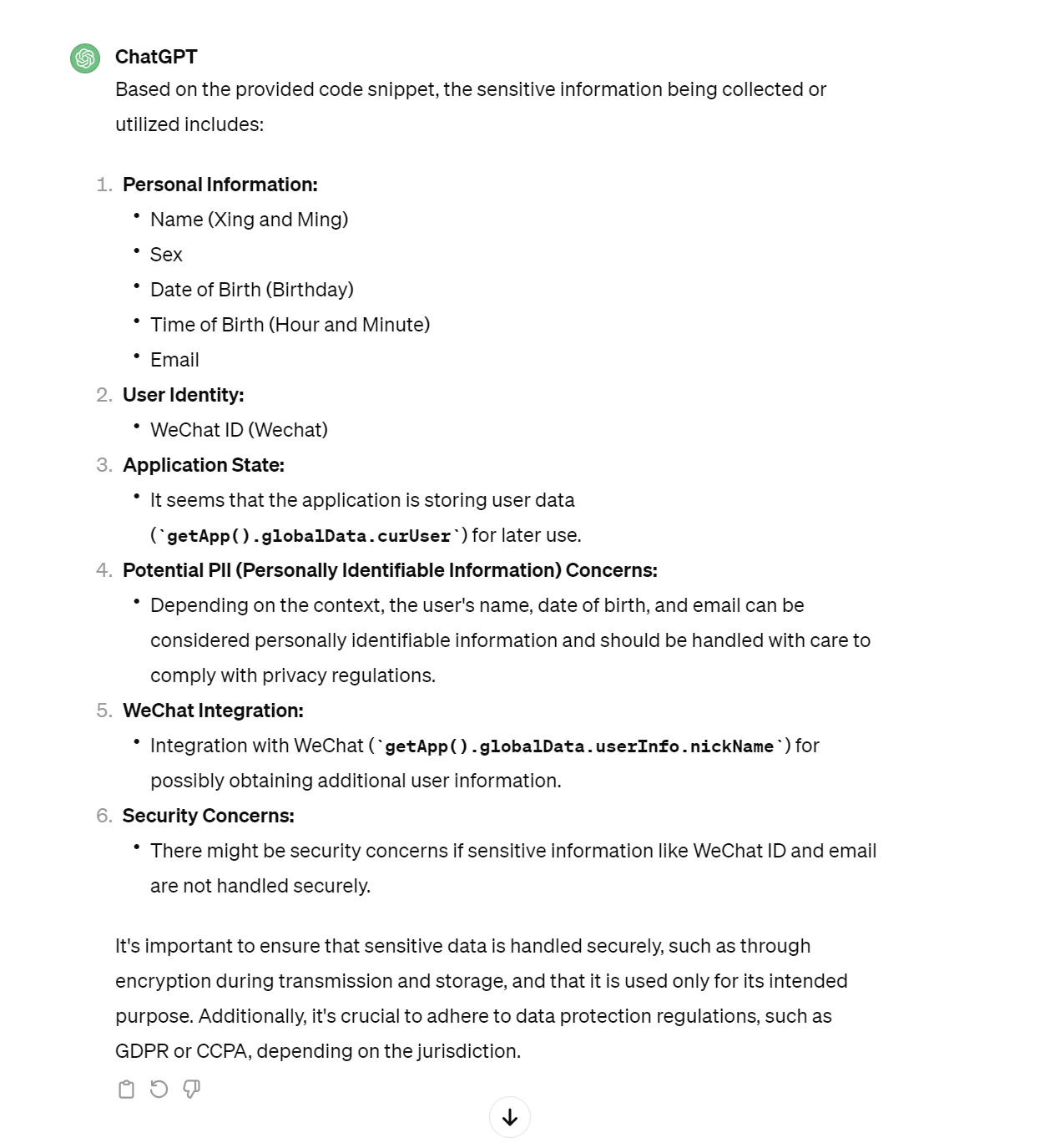}
    \caption{Caption}
    \label{fig:enter-label}
\end{figure}

\section{Discussion}
 
While LLMs offer advantages in their ability to understand context, process multimodal data, and scale to analyze large volumes of content, their effectiveness in detecting sensitive information within WeChat codes is subject to limitations such as language bias, complexity of content, privacy concerns, and the dynamic nature of communication on the platform. Addressing these limitations will be crucial in developing more accurate and reliable detection mechanisms for sensitive information within WeChat codes using LLMs. Frist, LLMs may exhibit biases based on the data they were trained on, which could lead to inaccuracies or inconsistencies in detecting sensitive information, particularly in cross-cultural contexts like WeChat, which is predominantly used in China. Certain culturally specific nuances or slang may be challenging for the model to interpret accurately. Second, WeChat codes can contain a wide range of content, including encrypted messages, multimedia files, and emoticons. Detecting sensitive information within such complex and varied content poses a challenge, as LLMs may struggle to interpret certain types of data accurately. Third, WeChat content is dynamic and constantly evolving. New forms of communication, emojis, slang, and cultural references emerge over time, making it difficult for LLMs to keep up with and accurately interpret all variations of content.

\section{Related Work}

Numerous works and studies have meticulously explored the multifaceted landscape of Large Language Models (LLMs), providing in-depth analyses and insights from various perspectives. These comprehensive investigations serve as valuable resources, shedding light on the evolution, taxonomy, and applications of LLMs across diverse domains. 
One notable aspect covered in these work is the evolutionary trajectory of LLMs, tracing their development from early models to the state-of-the-art architectures seen today. These studies delve into the historical context, highlighting key milestones, breakthroughs, and advancements that have propelled the field forward. Understanding the evolution of LLMs provides crucial context for assessing their current capabilities and anticipating future developments~\cite{yao2023survey,chang2023survey,zhao2023survey, wu2023survey,hadi2023survey,wu2023unveiling, bowman2023eight,zhao2023knnicl,fan2023large,hou2023large,thirunavukarasu2023large,clusmann2023future}.

The first study on miniapp security traces back to Lu et al.\cite{lu2020demystifying}, who investigated the security management of resources within app-in-app systems. In 2021, Zhang et al.\cite{ourfancywork} introduced MiniCrawler, assessing various aspects such as resource usage and API utilization. Wang et al.\cite{wang2022characterizing} proposed WeDetector in 2022, targeting prevalent bug patterns in WeChat mini-programs. Zhang et al.\cite{zhang2022identity} examined identity confusion vulnerabilities in app-in-app ecosystems, introducing CMRF~\cite{cmrf} and CMRFScanner. Zhang et al.\cite{zhang2023trusteddomain} revealed the Trusted Domain Compromise Attack, while Cai et al.\cite{cai2023shared} discussed user-centric account security. Wang et al.\cite{wang2023towards} compared traditional browser and super-app threat models, and Zhao et al.\cite{zhao2023potential} explored signature verification for enhancing miniapp ecosystem security. Zhang et al. unveiled vulnerabilities related to master key leakage~\cite{appsecretleak} and introduced TaintMini~\cite{taintmini2023} for tracking sensitive data flow. They also discovered hidden APIs~\cite{hiddenAPI2023} and revealed API execution variations across platforms with APIDiff~\cite{crossplatform2023}, focusing on user privileges and process isolation.

\section{Conclusion}

This paper delves into the potential of using Large Language Models (LLMs) to detect privacy breaches within WeChat Mini Programs. Given the ubiquitous use of Mini Programs and the increasing concerns surrounding data privacy, this research endeavors to assess the effectiveness of LLMs in identifying instances of privacy leakage within this ecosystem. Through thorough analysis and experimentation, we seek to shed light on the capability of LLMs to safeguard user privacy and security within the WeChat Mini Program environment. By contributing insights into enhancing privacy protection mechanisms, this research aims to foster a more secure and trustworthy digital environment for users.

\printbibliography

\end{document}